\documentstyle[sprocl,epsf]{article}
\begin{document}
%\draft

\title{COLLECTIVITY IN THE BRAIN SENSORY RESPONSE}

\author{{\bf S. Dro\.zd\.z}$^{1}$, J. Kwapie\'n$^{1}$, A.A.
Ioannides$^{2,3,4}$ 
and L.C. Liu$^{3}$}

\address{
$^{1}$ Laboratory for Nonlinear Dynamics, Institute of Nuclear Physics, 
PL--31-342 Krak\'ow, Poland,\\
$^{2}$ Laboratory for Human Brain Dynamics,
BSI, The RIKEN Institute, Hirosawa 2-1, Wako-shi 351-01, Japan,\\
$^{3}$ Institut f\"ur Medizin, Forschungszentrum J\"ulich,
D--52425 J\"ulich, Germany,\\
$^{4}$ Physics Department,  The Open University, Milton Keynes, MK7 6AA, UK}

%\date{\today}
\maketitle\abstracts{
A question of cooperative effects in auditory brain processing on various
space- and time-scales is addressed. The experimental part of 
our study is based on Multichannel Magnetoencephalography recordings 
in normal human subjects. Left, right and binaural stimulations were
used, in separate runs, for each subject.  The resulting time-series
representing left and right auditory cortex activity provide a clear evidence
for two levels of neuronal cooperation. One is the local hemispheric collective
response, termed M100 for its maximum at around 100ms after a stimulus onset.
Its only global characteristics turn out to be time-locked to a stimulus,
however, which means that the detailed neuronal evolution is largely
stochastic. This, together with the $1/f$ character of the corresponding 
power spectrum indicates self-organized criticality as an underlying 
mechanism. The second level is communication between the two hemispheres
with clear laterality effects: as a rule, the contralateral hemisphere leads
by $\sim$10ms. Mutual information analysis points to a direct information
transport as a way of this communication.} 

\section{Introduction}

The mammalian brain is the most complex system known to exist in the universe
and, for many reasons, for us human beings, it constitutes the most 
important organized structure. Understanding the principles of its organization
emerges the utmost intellectual challenge, also from the physics point of view.
Current understanding of complex systems indicates collectivity and chaos
as their two most basic and equally important characteristics.   
Even more, the most creative biological phenomena are believed to 
balance at the interface of collectivity and chaos~\cite{Kauf}. 
It seems natural to describe the brain in these terms, particularly its 
tendency to diversity and its ability of generating coherent patterns of 
activity, switching continuously from one to another. 
Such a view on the brain dynamics may also provide an appropriate frame
towards understanding this fundamental aspect of cortical organization
that manifests itself in local specialization and functional global 
integration~\cite{Fris}, both occuring on various space and time scales. 
A quantitative exploration of such a coexistence of space-time structures 
is of crucial importance for a proper design of a unified theoretical model 
relating local neuronal dynamics and global attributes of sensory processing.
With this in mind we have studied the activity in the human auditory cortex 
in response to simple tones, delivered regularly to one or both ears. 
Although even in this simple scenario animal~\cite{Aitk} and 
human studies~\cite{Laut} have shown that many 
different areas are involved, for this case the two auditory
areas are known to be active and prominent.

It is generally accepted that the physiology of the cortex 
is highly uniform~\cite{Vali}.
Experiments, for instance, indicate~\cite{Sur} that if the connections 
from the sensory organs to the auditory and visual areas of certain young
mammals are interchanged, significant functionality is still retained.
This suggests that the differences in function amoung the various cortex 
areas originate more from their different connections rather then from 
their distinct intrinsic characteristics. For this reason certain conclusions
to be drawn from the present study, focused on the auditory response,
may apply to the other sensory modalities as well.      

Information transfer within the brain is associated with weak electric currents
which results in electric potential and magnetic field.
When such currents in many nearby neuronal cells act in synchrony, 
the potential and fields grow large enough to be detected outside of the
skull
and the corresponding techniques are known as Electroecephalography (EEG)
and Magnetoencephalography (MEG)~\cite{Hama}. 
MEG is particularly appropriate in the present context, because 
activity from the auditory cortex is readily identifiable from both the 
average MEG signal~\cite{Hari,Sing} and in single trials~\cite{Liu1,Liu2}

MEG is a completely non-invasive method of measuring the distribution 
and time dependence of the magnetic field outside the skull.  
Similarly as the more conventional EEG it allows to
time-resolve neuronal activity down to the scale of 1ms~\cite{Creu}. Its main 
advantage over  scalp-EEG is that the skull and the scalp are transparent to 
the magnetic field and, therefore, an external measured magnetic field is
not distorted by radial conductivity effects. Furthermore, magnetic fields
outside the skull are generated predominantly by the currents tangential
to the surface of the head. The cortical currents are perpendicular 
to the surface of the cortex but almost $70\%$ of the human cortex is folded
into fissures which makes these currents effectively tangential to the skull
and, thus, accessible to MEG.
The above aspects of MEG make it particularly suitable 
for studying the spatio-temporal characteristics of the brain dynamics.

\section{MEG experiment}

The measurement of the magnetic field generated by the cortical 
neurons can be recorded   
using super Quantum Interference Devices (SQUIDs) operating within shielded 
environment~\cite{Hama}.   
In this contribution we will report a study performed with the twin MAGNES 
system of Biomagnetic Technologies inc. (BTi) in San Diego.  
This system has two separate dewars each with 37 first order gradiometers. 
During the experiment, the subject's head was resting on the bottom dewar, 
while the top dewar was placed over the opposite temporal area.  
Fig.~1 schematically illustrates this MEG setup.

Five healthy male volunteers (JD, JL, BD, FB and RB, age: 37.8$\pm$9.7) 
participated in the experiment.  
Four subjects (JD, JL, FB and RB) were right handed, 
two of them (FB and RB) were twins and one subject (DB) was left handed.
The stimuli were 50 msec, 1 kHz tone bursts at 
50dB (10 msec rise/fall and 30 msec plateau). The inter-stimulus interval 
was 1 second  ($\pm$ 20 msec). 
The MEG signal was recorded in continuous mode, 
sampled at 1042 Hz and filtered in real time with 0.1 Hz high pass. 
The analysis to be reported in this paper used two more signals obtained by 
further band-pass filtering in the 1-200 Hz (with notch filters at 50 Hz, 
100 Hz and 150 Hz),  and 3-20 Hz.
The subjects position was fixed as follows:  
A standard auditory evoked response was first obtained from stimuli 
delivered to both ears. This response is termed M100,  
it is the magnetic analogue of the N100,  a peak in the
EEG signal corresponding to the crest of a negative potential~\cite{Creu}.
The inspection of the average signal over 120 single trials
was used to guide repositioning of the dewars so that the
prominent M100 peak was captured with the positive and negative fields evenly
covered by the sensors in each probe. The procedure was repeated until 
each dewar was well positioned,  usually in one to three placements. 
Two further runs were obtained with this optimal dewar position with 
exactly the same protocol,  but with the stimulus delivered first to the 
left and then to the right ear.  The main experiment consisted of 
three runs:  the last dewar placement run with binaural stimuli and the 
two monaural stimulations.     

With optimal sensor location,  a very simple 
linear combinations of signals can be established to map the activity in 
each auditory cortex.  In effect we make from each 37 channel sensor array 
a Virtual Sensor (VS) which registers the activity 
in the adjacent auditory cortex~\cite{Liu2}.
For each probe, we have identified the two channels ($k_1$ and $k_2$), 
which produced the maximum difference at the 
time of the M100 peak, and used them to define the composite VS, 
\begin{equation}
VS^{M100} (t) =  \sum_{j=1}^{37} \, 
\left[ e^{-\left(\frac{|{\bf r}_j-{\bf r}_{k_1}|}{\lambda}\right)^2 } -  
e^{-\left(\frac{|{\bf r}_j-{\bf r}_{k_2}|}{\lambda}\right)^2 } \right]  
S_j (t) 
\end{equation}
where $\lambda$ is the characteristic length (we have used $\lambda$ = 0.02 m 
which is roughly the inter-channel separation);  
the results do not depend critically on this value.    
$S_j (t)$ is the MEG signal at time $t$ recorded by the $j^{th}$ channel, 
whose position vector is ${\bf r}_j$.
The coefficients of the expansion are computed at the time of the M100 peak 
in the average signal;  these coefficients are used unchanged for the analysis 
of all single trials.  

Fig.~2 shows a typical set of the average MEG signals corresponding to the
sensor arrangement as shown in Fig~1.

\section{Auditory hemispheric response}

From here on we will restrict our attention to the VS output generated as 
described above. Each run is then represented by two sets of the time 
series covering the 1s long time-interval $x^{\alpha}_L(t_n)$ and
$x^{\alpha}_R(t_n)$  ($n=1,...,1042$, corresponding
to the left (L) and the right (R) hemisphere, respectively.   
The sampling rate is 1042 Hz,  so $t_{n+1}-t_n=0.96$ms).
The superscript $\alpha=1,...,120$ labels the single trials in each experiment.
The time-series are consistently centered such that the onset of the stimulus 
corresponds to $n=230$.
Fig.~3 shows three typical, randomly selected, single-trial raw 
time-series together with the average 
\begin{equation}
x_{L,R}(t_n)= {1 \over N} \sum_{\alpha} x_{L,R}^{\alpha}(t_n)
\label{eq:ave}
\end{equation}
over all $N=120$ trials for the left (L) and right (R) hemisphere signals, for 
two subjects: JD and FB.   

It appears difficut to identify the stimulus onset from the raw single-trial
signal, although a relationship between the peak of the average response 
can be seen in some of the single trials.  
This reflects the fact that the conscious human brain is never at rest; 
central control of body function and regulation,  
fleeting thoughts and feelings, 
ensure that even in the most relaxed state a tapestry of regional activations 
is woven every instant.  Even the simplest of acts engages a multitude of 
areas in a way that varies even as the same task is repeated many times.
This explains why the single trial activity is not dominated by the stimulus. 
Even the same signals filtered to the frequency window of 3-20 Hz, 
as shown in Fig.~4, do not convincingly reflect the stimulus onset.
This indicates that the background brain activity may extend 
over the full spectrum of frequencies.   

However, since such a background brain activity is not time-locked 
to the stimulus it is averaged out after summing
up a sufficiently large number of identical trials.    
The average over the full set of our 120 consecutive trials exhibits a 
pronounced M100 peak centered at around 100ms (see Fig.~3) 
after the stimulus onset.
This peak reflects collective neuronal action at the superficial part of 
the auditory cortex in response to the stimulus.
The amplitude of variation of the corresponding field may, however, differ
from subject to subject as the two examples in Fig.~3 and ~4 indicate.
This signals differences in the degree of collectivity in the underlying
neuronal dynamics. A general observation is that this degree of
collectivity is significantly subject dependent but remarkably stable for 
each subject. This is more systematically documented in Fig.~5 which shows 
the avaraged (over 120 trials) MEG time-series for JD, DB, JL and FB
(RB looks very similar to FB). The result of both, 
the left and the right hemisphere responses generated 
by the left, right and binaural stimulations is here shown.

An interesting observation about the nature of such local auditory excitations
can be made by inspecting the structure of the power spectrum
calculated as a squared modulus of the Fourier transform of the time-series:
\begin{equation}
X_{L,R}(k) = \sum_{n=1}^N x_{L,R}(t_n) \exp(2 \pi i n k/N),
\label{eq:ft}
\end{equation}     
$X_{L,R}(k)$ being the complex numbers
$(X_{L,R}(k) = \vert X_{L,R}(k) \vert \exp(i\eta (k))$.
The power spectra are calculated from the time-series $x_{L,R}(t)$
representing the whole specific experiment lasting 120s and are 
shown in Fig.~6 for all five subjects participating in the experiment.
These power spectra somewhat differ amoung the subjects. The main difference
however is that JD (and to a lesser degree JL) exhibits a significant 
concentraction of strength at around 8 Hz but this can entirely be attributed
to a particularly strong $\alpha$-rythm activity dominating this subject.
Ignoring this peak one obtains a very similar
"$1/f$"-type (straight line of finite negative slope
in the log-log scale) global behavior~\cite{Dutt}.
This may be considered as an indication that evolution 
of the M100 complex is governed by a very universal phenomenon of
self-organized criticality~\cite{Bak} which is a more
catastrophic form of collectivity and is generated by a fractal 
(scale-invariant) 'avalanche'-like process.
Interestingly, a new class of neural networks based on {\sl adaptive
performance networks}~\cite{Alst} shows exactly this type of power spectra.
It also allows some local deviations from this behavior and those deviations
result from certain subject specific stronger activity at some frequency.
This model involving the elements of self-organized criticality can be
trained~\cite{Stas}  to react 'intelligently' to external sensory signals.
It is also interesting to notice that the power indices corresponding to
different subjects are not exactly the same. 
Relating these indices listed in Fig.~6 with signal amplitudes 
at M100 displayed in Fig.~5 shows that the cases of stronger collectivity
(JD) are accompanied by the power spectra whose slope is 
somewhat amplified relative to the cases of weaker collectivity (FB, RB). 
Larger slope means stronger deviation from the pure white noise phenomena.
As it is thus natural, in this case the weaker collectivity is connected 
with a more noisy dynamics which acts destructively on local coherence.

Chaotic versus regular dynamics can be quantified in terms of the Lyapunov
exponents $\lambda$ which characterize a degree of divergence or convergence of
nearby orbits in phase space. A numerical procedure~\cite{Wolf}
of calculating them directly from the time-series $x(t)$ is based on an 
m-dimensional phase portrait whose points on the attractor 
are given by $\{x(t),x(t+\tau),...,x(t+[m-1]\tau)\}$,
where $\tau$ denotes an appropriately chosen delay time.
To the initial point at $t=t_0$ one then locates its nearest neighbor.
The distance $L^0(t_0)$ between such two points evolves to $L(t_1)$
at a latter time $t_1=t_0+\Delta$. $\Delta$ needs to be sufficiently small
so that only small scale attractor structure is examined. In practice,
one therefore $M$-times repeats the above procedure until the entire data 
file is traversed and one estimates
\begin{equation}
\lambda={1\over{M\Delta}} \sum_{k=1}^M \log {L(t_k)\over{L^0(t_{k-1})}}.
\label{eq:lambda}
\end{equation}    
Optimal value for the delay time $\tau$ is the one which corresponds
to the minimum in the correlation function of $x(t)$.  
For our MEG time-series it consistently results in values 30-35ms. 
When estimating the maximum value of $\lambda$, which is our purpose here,
its value does not critically depend on $m$. It only needs to be sufficiently
large. In our case $m=4$ fulfils this requirement.
Fig.~7 shows the $\Delta$ dependence of $\lambda$ as defined by 
eq.~\ref{eq:lambda} for the same two subjects whose MEG time-series are shown
in Figs.~3 and ~4, i.e., for the raw as well as for the filtered signals. 

Consistent with a more noisy character of the time-series for the subject
FB, the corresponding $\lambda(\Delta)$ for this subject is almost an order of
magnitute larger than for JD (note different scales in Fig.~7). 
This observation applies on the level of both, 
the raw and the filtered signals. For both subjects the raw 
data do not lead to any plateau in $\lambda(\Delta)$. 
This kind of a behavior is characteristic for complex multidimensional 
dynamical systems. Such systems are, in fact, indistinguishable from noisy.  
Our filtered time-series, however, do already display a significant stability
over relatively large intervals of $\Delta$ and, thus, the corresponding
values of $\lambda$ can be considered the characteristic largest Lyapunov
exponents. In this case we are thus dealing with a low dimensional 
deterministic chaos.

\section{Interhemispheric correlations}

As documented above, the MEG signals carry signatures of strongly chaotic
systems. Still, one may expect some sort of 
long-range interhemispheric correlations.
This can be anticipated from the structure of average signals shown in
Fig.~5. Even though the stimulus is applied asymmetrically (left or right 
ear) a similar (but not identical) structure is detected on both hemispheres. 
A question of primary interest is what is the nature and what is the mechanism 
of this effect. Is it an independent activation of both auditory areas
seen only in the average signal, or is it a genuine information transport
from one area to the other? 

An appropriate theoretical tool to quantify such 
effects is provided by the concept of mutual information~\cite{Swin}.
It evaluates the amount of information about one of the subsystems $(s1)$
resulting from a measurement of the other $(s2)$ and is defined as 
\begin{equation}
I(X_{s1},X_{s2})=H(X_{s1}) + H(X_{s2}) - H(X_{s1},X_{s2}).
\label{eq:mi}
\end{equation}
$X_s$ denotes here the whole set of possible messages about the subsystem $s$
and $H(X_s)= - \sum_j p(j) \ln p(j)$ is the corresponding entropy evaluated
from probabilities $p(j)$ that $x_s(t)$ assumes value characteristic for 
$j$th element of the partition. $H(X_{s1},X_{s2})$ is a joint entropy for
the combined system, calculated analogously from the joint probability
$p(j_1,j_2)$.
It easy to verify that
\begin{equation}
I(X_{s1},X_{s2}) \ge 0
\label{eq:ge}
\end{equation}
and the equality holds only if $s1$ and $s2$ are statistically independent,
i.e., $p(j_1,j_2)=p(j_1)p(j_2)$.

Information transport between the subsystems may lead to time-delayed 
effects in the synchronization of correlations. 
Such effects can be evaluated by calculating the time-delayed mutual 
information between measurements of the two subsystems at different times. 
The corresponding prescription retains the structure of Eq.~\ref{eq:mi}.
The time-series $x_{s1}(t)$
needs only to be correlated with $x_{s2}(t+\tau)$. The mutual information
$I(X_{s1},X_{s2};\tau)$ then becomes a function of the time-delay $\tau$.     
It may display maximum at a certain finite value of $\tau$. This value of 
$\tau$ thus provides an estimate on the time needed for the information to be
transported from the subsystem $s1$ to $s2$.  

For the present application we find~\cite{Kwap} it more useful to apply 
the generalized entropy ~\cite{Reny}:
\begin{equation}
H_q(X_s)= {1 \over {1-q}} \ln \sum_j p^q(j)
\label{eq:gent}
\end{equation}
For $q \to 1$ this equation yields the standard information entropy.
The most useful property of $H_q(X_s)$ is that with
increasing $q$ a higher weight is given to the largest components in the
set $\{p(j)\}$. This proves very instructive in studying various aspects of the
phase-space exploration in dynamical systems~\cite{Droz}.
One then obtains~\cite{Kwap} the following expression for the generalized 
mutual information:
\begin{equation}
I_q (X_{s1},X_{s2};\tau) = {1 \over {1-q}} \ln {{\sum_{j_1} p^q(j_1) 
\sum_{j_2} p^q(j_2)} \over {\sum_{j_1j_2} p^q(j_1,j_2;\tau)}},
\label{eq:qmi}
\end{equation}
which constitutes a basis for our numerical applications.
Indeed, the higher $q$-values offer a much more precise estimate for the
time-delay $\tau$ at maximum and this is especially important for weak
correlations. $q=6$ we find~\cite{Kwap} satisfactory for analysing the
present data. In the following, when making use of this equation,
a grid of 10 bins covering an interval of variation of both,
$x^{\alpha}_R(t_n)$ and $x^{\alpha}_L(t_n+\tau)$ is introduced. This guarantees
stability of the results. For a given experiment the three different 
probability distributions entering Eq.~\ref{eq:qmi} are evaluated by 
superimposing histograms corresponding to all the time-series 
$(\alpha=1,...,120)$ and then the logarithm is taken.
A convention used in the corresponding calculation when defining
the sign of the time-delay $\tau$ between $x_L(t_n)$ and $x_R(t_n+\tau)$ 
is such that its negative value means that a relevant excitation in the right 
hemisphere is time-advanced relative to the left hemisphere. Of course, the
opposite applies for positive sign.                      

An important related issue is localization of correlations in frequency.
Therefore, we first explore the variation of $I_q(\tau)$ $(q=6)$
between the two hemispheres as a function of the frequency. 
By inverting the discret Fourier transform (Eq.~\ref{eq:ft}) of the input 
data series $x_{L,R}(t_n)$ in a reduced interval 
$\langle K- \Delta K/2,K+ \Delta K/2 \rangle$ of discrete frequencies $k$
one obtains the filtered series $x_{L,R}^{K,\Delta K}(t_n)$
spanning the frequency window $\Delta K$ centered at $K$: 
\begin{equation}
x_{L,R}^{K,\Delta K}(t_n) = {1 \over \Delta K}
\sum_{k=K-\Delta K/2}^{K+\Delta K/2} X_{L,R}(k) \exp(-2 \pi i n k/N).
\label{eq:wind} 
\end{equation}

Fig.~8 shows a typical landscape of the mutual information in the
time-delay
$\tau$ and in the frequency window of 5 Hz (which is sufficiently large 
so that no artificial correlations are generated~\cite{Kwap})
centered at the value indicated.
As it is clearly seen, the correlations exist and  
are mediated by the low-frequency (up to 20 Hz) activity.        
This picture turns out to be subject independent. The amplitude of MI is
found to depend from subject to subject, however. For certain subjects the
correlations are much weaker and they would be hardly identifiable on the level
of $q=1$ MI.  

Another question of principal interest is how do the correlations relate
to the stimulus onset. Our method of quantifying correlations allows to
study them for relatively short time-series, even taking the 100ms long 
time-intervals of our original series. Fig.~9 shows a similar landscape of
MI as Fig.~8 but now in $\tau$ versus the time window of 100ms 
and the whole spectrum of frequencies is covered.   

Fig.~9 thus indicates that the appearence of correlations is connected 
mainly with the M100 peak.
Thus, for a more systematic study of $I_q(\tau)$ the time-series will be
truncated to the interval between $i=230$ and $i=491$. This covers 250ms
starting exactly at the initial moment of the stimuls. Furthermore,
according to the above frequency localization, and in order to make this
study more transparent, all the time-series are filtered to the frequency
window between 3 and 20Hz.    
The results for four subjects (RB looks similar to FB) are collected in
Fig.~10. Several conclusions are to be drawn from this figure.
First of all, the correlations under study are spatially nonuniform
and the information transport between the hemispheres takes about 10ms.
The relative location of the peaks in MI indicates that, at least
statistically,  the contralateral hemisphere drives the response
for all the subjects and conditions studied. This, however, in general 
can only be identified by a parallel analysis of the left versus right ear 
stimulation (binaural is also helpful) of the same subject. 
The point is that for some subjects there are certain asymmetry effects. 
For instance, in JL the ipsilateral hemisphere somewhat overtakes ($\sim$ 5ms)
when the right ear is stimulated but then the contralateral hemisphere 
overtakes even more when the tone is delivered to the other ear, 
so that the relative location of the peaks in MI, corresponding to the 
left and right ear stimulation, respectively, is still preserved.
A trace of asymmetry, but in opposite direction, is also visible in JD.
A likely explanation of those asymmetry effects
is that we are facing a superposition of the two phenomena. 
One is a leading role of the contralateral hemisphere when the tone is
delivered to one ear (either left or right) and the other may originate from
certain subject specific asymmetry in properties of the left and right auditory
areas. The latter kind of asymmetry is known to occur quite
frequently~\cite{Creu}.

A related quantity of interest is the strength of information transfer 
between the hemispheres. 
This characteristics measured in terms of the MI-excess over background 
is largely invariant for a given subject (similar for different experiments). 
It is, however, strongly subject dependent and ranges between very
pronounced (JD) and rather weak (FB, RB). This effect may in principle 
originate from two different sources.
One is different strength of the coupling between the hemispheres, the other
is subject dependent degree of collectivity of the local M100 excitations.
That the second possibility is more likely to apply here can be concluded from
Fig.~5. The amplitude of the average signal at M100 
goes in parallel with the strength of the information transport.
This, in fact, is natural since the amount of information to be communicated
results from the original local collectivity.
It is also consistent with the low-frequency origin of
inter-hemispheric correlations as illustrated in Fig.~8. Localization in
frequency means higher synchrony and more determinism, 
and these are expected to constitute preferential
conditions for the long-range inter-hemispheric correlations to occur.  

In order to understand the mechanism of inter-hemispheric 
correlations, it is instructive to look at MI between $x^{\alpha}_L(t)$
and $x^{\alpha + \delta}_R(t)$ for $\delta \ne 0$.
Fig.~11 convincingly documents that such correlations are much weaker for 
both subjects shown. This holds true for all the remaining subjects as well.
This is a very interesting and meaningful result which, first of all, indicates
that different configurations of neurons are excited in each trial.     
In other words, the specific evolution of M100 with respect to
consecutive trials must involve nondeterministic elements which make the
above, translated correlations much weaker.
Consequently, only the global aspects of M100 are time-locked 
to the stimulus; a detailed 'microscopic' evolution turns out 
largely stochastic. Secondly, this result indicates 
that what actually correlates the opposite hemispheres
in the present context is not just an independent appearance of M100 in both
hemispheres but the real inter-hemispheric information transport which 
projects one M100 into another and thus induces certain similarity between 
them. They are thus functionally related and this is what the corresponding 
peaks in the mutual information reflect.

\section{Conclusions}

The study presented above provides a clear evidence for two levels 
of dynamical cooperation in the brain auditory processing, occuring on two
different scales.
One is the local hemispheric collective response, reaching its maximum at
about 100ms (M100) after a stimulus onset. 
An interesting emerging property of this excitation is that its 
only global characteristics are time-locked to a stimulus.
The underlying neuronal degrees of freedom involved are likely 
to significantly differ from trial to trial in a quite stochastic manner. 
In principle such a behavior may occur 
in certain rather standard neural network models~\cite{Trau}. 
However, a possible natural scenario potentially able to reconcile these 
two aspects of evolution of the M100 response 
and the inverse power-law character of the corresponding power spectrum 
is self-organized criticality~\cite{Bak}. 
The second level of cooperation is the long-range communication between 
the two hemispheres. The most conclusive in this connection are the monaural
stimulations. The analysis based on mutual information then shows that, 
at least statistically, the contralateral hemisphere 
systematically leads by 10-20ms. The mechanism
of this communication carries the signature of (delayed) synchronization
and thus can be hypothesised as a direct information transport
between the hemispheres. This process can also be considered as an example
of the most general effect of synchronization~\cite{Amen} in a spatially 
extended chaotic system.

\section*{Acknowledgments}

This work was supported in part by Polish KBN Grant No. 2 P03B 140 10.

\newpage

% --------------------------  Figure captions  -------------------------

{\bf Fig.~1}
Coronal and sagittal views showing the sensor arrangement relative
to the head and brain.

{\bf Fig.~2}
The average MEG signal for tone presentation to
the left ear in the channels of the left and right probe.  The channels
with the strongest positive and negative signal are marked for each probe.
The difference of weighted sums of channels, with weights decreasing with
distance away from the highlighted channels define the Virtual Sensor.

{\bf Fig.~3}
Three randomly selected raw MEG time-series (dashed, dash-dotted
and dotted lines) for the subject JD (left column) and FB (right column)
corresponding to left ear stimulation. Upper part illustrates the right
hemisphere and lower part the left hemisphere behavior.

{\bf Fig.~4}
The same as in Fig.~3 corresponding to left hemisphere, but the
MEG signal is now filtered to the 3-20 Hz frequency window.

{\bf Fig.~5}
The avaraged MEG time-series over all 120 trials for four
different subjects corresponding to the left ear (LE), right ear (RE)
and binaural (B) stimulation.
The solid line displays the left hemisphere and the dashed line the right
hemisphere response.

{\bf Fig.~6}
Typical power spectra of the full MEG time-series for all five subjects.

{\bf Fig.~7}
$\Delta$ dependence of $\lambda$ as defined by
eq.~\ref{eq:lambda} for JD and FB. Solid line corresponds to the raw data
while the dashed line describes the MEG time-series filtered to the 3-20
Hz frequency window.

{\bf Fig.~8}
$I_6(\tau)$ as a function of the frequency
(frequency window of 5 Hz) for subject JD, left ear stimulation.

{\bf Fig.~9}
$I_6(\tau)$ as a function of the time-window of 100ms
for subject JD, left ear stimulation. Zero corresponds to the stimulus
onset.

{\bf Fig.~10}
$I_6(\tau)$ as a function of the time-delay for all five subjects
calculated from the time-interval between 0 (stimulus onset) and 250ms.
Solid line displays the response to the left ear, dashed line to the right
ear and dash-dotted line to the binaural stimulation.

{\bf Fig.~11}
Two examples (for JD and FB) of $I_6(\tau)$ between the time-series
representing different trials, i.e., $x^{\alpha}_L(t)$ is correlated with
$x^{\alpha + \delta}_R(t)$. The soild line corresponds to $\delta=0$
(original case), $\delta=1$ to the dotted line, $\delta=4$ to the
dash-dotted line and $\delta=10$ to the dashed line.

\end{document}